\documentstyle[11pt,amssymb]{article}
\let\a=\alpha    
    
  \let\n=\nu

\let\C=\Chi

\def\nn{\nonumber} \def\bd{\begin{document}} \def\ed{\end{document}}
\def\ds{\documentstyle} \let\fr=\frac \let\bl=\bigl \let\br=\bigr
\let\Br=\Bigr \let\Bl=\Bigl 
\let\bm=\bibitem
\let\na=\nabla
\let\pa=\partial \let\ov=\overline 
\newcommand{\be}{\begin{equation}} 
\newcommand{\ee}{\end{equation}} 
\def\ba{\begin{array}}
\def\ea{\end{array}}
\def\ft#1#2{{\textstyle{{\scriptstyle #1}\over {\scriptstyle #2}}}}
\def\fft#1#2{{#1 \over #2}}
\def\del{\partial}
\def\vp{\varphi}
\def\st#1{{\scriptstyle #1}}
\def\sst#1{{\scriptscriptstyle #1}}

\def\oneone{\rlap 1\mkern4mu{\rm l}}
\def\td{\tilde}
\def\wtd{\widetilde}
\def\ie{\rm i.e.\ }
\def\dalemb#1#2{{\vbox{\hrule height .#2pt
        \hbox{\vrule width.#2pt height#1pt \kern#1pt
                \vrule width.#2pt}
        \hrule height.#2pt}}}
\def\square{\mathord{\dalemb{6.8}{7}\hbox{\hskip1pt}}}

\def\cramp{\medmuskip = 2mu plus 1mu minus 2mu}
\def\cramper{\medmuskip = 2mu plus 1mu minus 2mu}
\def\crampest{\medmuskip = 1mu plus 1mu minus 1mu}
\def\uncramp{\medmuskip = 4mu plus 2mu minus 4mu}

\newcommand{\ho}[1]{$\, ^{#1}$}
\newcommand{\hoch}[1]{$\, ^{#1}$}
\newcommand{\bea}{\begin{eqnarray}} 
\newcommand{\eea}{\end{eqnarray}} 
\newcommand{\ra}{\rightarrow}
\newcommand{\lra}{\longrightarrow}
\newcommand{\Lra}{\Leftrightarrow}
\newcommand{\ap}{\alpha^\prime}
\newcommand{\bp}{\tilde \beta^\prime}
\newcommand{\tr}{{\rm tr} }
\newcommand{\Tr}{{\rm Tr} } 
\def\0{{\sst{(0)}}}
\def\1{{\sst{(1)}}}
\def\2{{\sst{(2)}}}
\def\3{{\sst{(3)}}}
\def\4{{\sst{(4)}}}
\def\5{{\sst{(5)}}}
\def\6{{\sst{(6)}}}
\def\7{{\sst{(7)}}}
\def\8{{\sst{(8)}}}
\def\n{{\sst{(n)}}}
\def\cA{{{\cal A}}}
\def\cF{{{\cal F}}}
\def\tV{\widetilde V}
\def\tW{\widetilde W}
\def\tH{\widetilde H}
\def\tE{\widetilde E}
\def\tF{\widetilde F}
\def\tA{\widetilde A}
\def\im{{{\rm i}}}
\def\jm{{{\rm j}}}
\def\km{{{\rm k}}}

\def\tY{{{\wtd Y}}}
\def\ep{{\epsilon}}
\def\vep{{\varepsilon}}
\def\R{\rlap{\rm I}\mkern3mu{\rm R}}
\def\bD{{{\bar D}}}
\def\R{{{\Bbb R}}}
\def\C{{{\Bbb C}}}
\def\H{{{\Bbb H}}}
\def\CP{{{\Bbb C}{\Bbb P}}}
\def\RP{{{\Bbb R}{\Bbb P}}}
\def\Z{{{\Bbb Z}}}
\def\bA{{{\Bbb A}}}
\def\bB{{{\Bbb B}}}

\newcommand{\NP}{Nucl. Phys. }
\newcommand{\tamphys}{\it Center for Theoretical Physics\\
Texas A\&M University, College Station, TX 77843, USA}
\newcommand{\umich}{\it Michigan Center for Theoretical Physics\\
University of Michigan, Ann Arbor, Michigan 48109, USA}
\newcommand{\upenn}{\it Department of Physics and Astronomy\\
University of Pennsylvania, Philadelphia,  PA 19104, USA}
\newcommand{\SISSA}{\it  SISSA-ISAS and INFN, Sezione di Trieste\\
Via Beirut 2-4, I-34013, Trieste, Italy}

\newcommand{\ihp}{\it Institut Henri Poincar\'e\\
  11 rue Pierre et Marie Curie, F 75231 Paris Cedex 05}

\newcommand{\damtp}{\it DAMTP, Centre for Mathematical Sciences,
 Cambridge University, Wilberforce Road, Cambridge CB3 OWA, UK}

\newcommand{\auth}{M. Cveti\v{c}\hoch{\dagger}, G.W. Gibbons\hoch{\sharp}, 
H. L\"u\hoch{\star\dagger} and C.N. Pope\hoch{\ddagger\dagger}}

\begin{document}
\begin{flushright}
\hfill{DAMTP-2001-25}\ \ \ {CTP TAMU-10/01}\ \ \ {UPR-931-T}\ \ \
{MCTP-01-14}\\ 
{March 2001}\ \ \
{hep-th/0103155}
\end{flushright}

%\vspace{15pt}

\begin{center}
{ \large {\bf New Complete Non-compact Spin(7) Manifolds}}

\vspace{8pt}
\auth

\vspace{5pt}
{\hoch{\dagger}\upenn}

\vspace{5pt}
{\hoch{\sharp}\damtp}

%%\vspace{5pt}
%%{\hoch{\dagger} \it Department of Physics and Astronomy, Rutgers University,
%%Piscataway, NJ 08855}

\vspace{4pt}
{\hoch{\star}\umich}

\vspace{4pt}
{\hoch{\ddagger}\tamphys}

%\vspace{5pt}
%{\hoch{\dagger,\sharp,\ddagger}\ihp}

\vspace{5pt}

\underline{ABSTRACT}
\end{center}

   We construct new explicit metrics on complete non-compact
Riemannian 8-manifolds with holonomy Spin(7). One manifold, which we
denote by $\bA_8$, is topologically $\R^8$ and another, which we
denote by $\bB_8$, is the bundle of chiral spinors over $S^4$.  Unlike
the previously-known complete non-compact metric of Spin(7) holonomy,
which was also defined on the bundle of chiral spinors over $S^4$,
our new metrics are asymptotically locally conical (ALC): near
infinity they approach a circle bundle with fibres of constant length
over a cone whose base is the squashed Einstein metric on $\CP^3$.  We
construct the covariantly-constant spinor and calibrating 4-form.  We
also obtain an $L^2$-normalisable harmonic 4-form for the $\bA_8$
manifold, and two such 4-forms (of opposite dualities) for the $\bB_8$
manifold.  We use the metrics to construct new supersymmetric brane
solutions in M-theory and string theory.  In particular, we construct
resolved fractional M2-branes involving the use of the $L^2$ harmonic
4-forms, and show that for each manifold there is a supersymmetric
example.  An intriguing feature of the new $\bA_8$ and $\bB_8$ Spin(7)
metrics is that they are actually the {\it same} local solution, with
the two different complete manifolds corresponding to taking the
radial coordinate to be either positive or negative.  We make a
comparison with the Taub-NUT and Taub-BOLT metrics, which
by contrast do not have special holonomy.  In an appendix we construct
the general solution of our first-order equations for Spin(7)
holonomy, and obtain further regular metrics that are
complete on manifolds $\bB_8^+$ and $\bB_8^-$ similar to
$\bB_8$.

%{\vfill\leftline{}\vfill
%\vskip 5pt
%\footnoterule
%{\footnotesize \hoch{1} Research supported in part by DOE grant
%DE-FG02-95ER40893 and NATO grant 976951. \vskip -12pt} \vskip 14pt
%{\footnotesize \hoch{2} Research supported in full by DOE grant
%DE-FG02-95ER40899 \vskip -12pt} \vskip 14pt
%{\footnotesize  \hoch{3} Research supported in part by DOE
%grant DE-FG03-95ER40917.\vskip  -12pt}}

%\baselineskip=24pt
\pagebreak
\setcounter{page}{1}

%\tableofcontents
\vfill\eject

\section{Introduction}

    There are many explicit examples of Ricci-flat metrics with
K\"ahler or hyper-K\"ahler special holonomy that are defined on
regular non-compact manifolds.  There are far fewer analogous examples
of Ricci-flat metrics with the exceptional holonomies $G_2$ in $D=7$
or Spin(7) in $D=8$.  In fact three explicit non-compact $G_2$
examples and one explicit Spin(7) example are known
\cite{brysal,gibpagpop}.  In this paper we obtain new
eight-dimensional metrics of Spin(7) holonomy, and show how they can
be defined on two topologically inequivalent regular non-compact
manifolds.  The new metrics are all asymptotically locally conical
(ALC), locally approaching $\R\times S^1\times \CP^3$.  The radius of
the $S^1$ is asymptotically constant, so the metric approaches an
$S^1$ bundle over a cone with base $\CP^3$.  However, the Einstein
metric on the $\CP^3$ at the base of the cone is not the Fubini-Study
metric, but instead the ``squashed'' metric described as an $S^2$
bundle over $S^4$.  The new solutions can have very different
short-distance behaviours, with one approaching flat $\R^8$ whilst the
others approach $\R^4\times S^4$ locally.  The global topology is that
of $\R^8$ in the first case and the bundle of positive (or negative)
chirality spinors over $S^4$ for the others.  An intriguing feature of
two of the new metrics, one on each of the inequivalent topologies, is
that locally they are actually the same.  This metric is complete on a
manifold of $\R^8$ topology if the radial coordinate is taken to be
positive, whilst in the region with negative $r$ it is instead
complete on the manifold ${\Bbb S}(S^4)$ of the bundle of chiral
spinors over $S^4$.  We shall denote the new Spin(7) manifold with
$\R^8$ topology by $\bA_8$, and the new related manifold with ${\Bbb
S}(S^4)$ topology by $\bB_8$.  In appendix A we construct the general
solution of the first-order equations that follow by requiring Spin(7)
holonomy in our metric ansatz, and we show that these lead to further
more general classes of regular metrics\footnote{This appendix with
the general solution and the further complete Spin(7) metrics extends
the results in an earlier version of this paper.} defined on complete
manifolds $\bB_8^\pm$ that are again topologically the bundle of
chiral spinors over $S^4$.

    Our construction is a generalisation of the one that leads to the
previously-known metric of Spin(7) holonomy.  That example is given 
by \cite{brysal,gibpagpop}
%%%%%
\be
ds_8^2 = \Big(1- \fft{\ell^{10/3}}{r^{10/3}}\Big)^{-1} \, dr^2 
         + \ft{9}{100}\, r^2\, \Big(1- \fft{\ell^{10/3}}{r^{10/3}}\Big)\,
         h_i^2 + \ft{9}{20} r^2\, d\Omega_4^2\,,\label{spin7metric}
\ee
%%%%%
where
%%%%%
\be
h_i\equiv \sigma_i - A_\1^i\,,
\ee
%%%%%
the $\sigma_i$ are left-invariant 1-forms on $SU(2)$, $d\Omega_4^2$ is
the metric on the unit 4-sphere, and $A_\1^i$ is the $SU(2)$
Yang-Mills instanton on $S^4$.  The $\sigma_i$ can be written in terms
of Euler angles as
%%%%%
\crampest
\be
\sigma_1 = \cos\psi\, d\theta + \sin\psi\, \sin\theta\,
d\varphi\,,\quad
\sigma_2 = -\sin\psi\, d\theta + \cos\psi\, \sin\theta\,
d\varphi\,,\quad
\sigma_3 = d\psi + \cos\theta\, d\varphi\,.\label{1forms}
\ee
\uncramp
%%%%%
The principal orbits are $S^7$, described as an $S^3$ bundle over $S^4$.
The solution (\ref{spin7metric}) is asymptotic to a cone over the
``squashed'' Einstein 7-sphere, and it approaches $\R^4\times S^4$
locally at short distance (\ie $r\approx\ell$).  Globally the manifold
has the same topology ${\Bbb S}(S^4)$, the bundle of chiral spinors
over $S^4$, as the new Spin(7) manifolds $\bB_8$ and $\bB_8^\pm$ 
that we obtain in this paper.

\section{Ansatz, Einstein equation and superpotential for Spin(7) metrics}

   The generalisation that we shall consider involves allowing the
$S^3$ fibres of the previous construction themselves to be
``squashed.''  In particular, this encompasses the possibility of
having an asymptotic structure of the ``Taub-NUT type,'' in which the
$U(1)$ fibres in a description of $S^3$ as a $U(1)$ bundle over $S^2$
approach constant length while the radius of the $S^2$ grows linearly.
The appropriate squashing along the $U(1)$ fibres can be implemented
using a description given in \cite{clp0}, where it was observed that
if one defines
%%%%%
\be
\mu_1 = \sin\theta\, \sin\psi\,,\qquad 
\mu_2= \sin\theta\, \cos\psi\,,\qquad
\mu_3= \cos\theta\,,
\ee
%%%%%
then $h_i$ can be written (after adapting some conventions) as
%%%%%
\be
h_i = -\ep_{ijk}\, \mu^j\, D\mu^k + \mu^i\, \sigma\,,
\ee
%%%%%
where
%%%%%
\be
D\mu^i\equiv d\mu^i +\ep_{ijk}\, A_\1^j\, \mu^k\,,\quad
\sigma \equiv d\varphi + \cA_\1\,,\quad \cA_\1 \equiv \cos\theta\, d\psi -
\mu^i\, A_\1^i\,.\label{kkvector}
\ee
%%%%% 
It also follows that
%%%%%
\be
\sum_i h_i^2 = \sum_i (D\mu^i)^2 + \sigma^2\,,
\label{hexp0}
\ee
%%%%%
In terms of the coordinates $(\theta,\psi)$ on $S^2$, we have
%%%%%
\bea
\sum_i (D\mu^i)^2 &=& (d\theta - A_\1^1\, \cos\psi + A_\1^2\, \sin\psi)^2 
\nn\\
&& + \sin^2\theta\, (d\psi + A_\1^1\, \cot\theta\, \sin\psi + A_\1^2 \,
   \cot\theta\, \cos\psi - A_\1^3)^2\,.
\eea
%%%%%

  Finally, one can show that the field strength
$\cF_\2=d\cA_\1$, which follows from (\ref{kkvector}), is given by
%%%%%
\be
\cF_\2 = \ft1{2} \ep_{ijk}\, \mu^k\, D\mu^i \wedge
D\mu^j - \mu^i\, F_\2^{i}\,.\label{2f}
\ee
%%%%%
Since $\mu^i\, \mu^i=1$, we see that (\ref{hexp0}) expresses the
metric on the $S^3$ fibres as a $U(1)$ bundle over $S^2$, with fibre
coordinate $\varphi$.  Note that $\varphi$ has period $4\pi$, while
$\psi$ has period $2\pi$.  This reversal of the periods by comparison
to those for the left-invariant 1-forms (\ref{1forms}) is associated
with the fact that we effectively transformed from a left-invariant
basis to a right-invariant one, in passing to the metric (\ref{hexp0})
on $S^3$ \cite{clp0}.  The same transformation, expressed somewhat
differently, was used recently in \cite{3wm}.

    With these preliminaries, we can now present our more general ansatz for
8-dimensional metrics of Spin(7) holonomy:
%%%%%
\be
d\hat s_8^2 = dt^2 + a^2\, (D\mu^i)^2 + b^2\, \sigma^2 + c^2\,
d\Omega_4^2\,.\label{8ans}
\ee
%%%%%
Here $a$, $b$ and $c$ are functions of the radial variable $t$.  The
metric has cohomogeneity one, with principal orbits that are
homogeneously-squashed $S^7$.  The previous Spin(7) example
(\ref{spin7metric}) has $a=b$.

   A convenient way to obtain the conditions for Ricci-flatness for the ansatz
(\ref{8ans}) is to perform a Kaluza-Klein dimensional reduction on the
$U(1)$ fibres parameterised by the $\varphi$ coordinate.  This reduction
can be written as 
%%%%%
\be
d\hat s_8^2 = e^{-\fft1{\sqrt{15}}\phi}\, ds_7^2 +
e^{\sqrt{\fft53}\phi}\, (d\varphi+ B_\1)^2\,,\label{kkred}
\ee
%%%%%
where $ds_7^2$, $B_\1$ and $\phi$ are all independent of the fibre 
coordinate $\varphi$.  The conditions for Ricci flatness of the 
eight-dimensional metric (\ref{kkred}) are then equivalent to the 
seven-dimensional Einstein-Maxwell-Dilaton equations that follow from
the dimensional reduction of the Einstein-Hilbert Lagrangian, which 
in $D=7$ gives
%%%%%
\be
e^{-1}\, {\cal L}_7 =R - \ft12(\del\phi)^2 -\ft14
e^{2\sqrt{\fft35}\,\phi}\, G_\2^2\,,
\ee
%%%%%
where $G_\2=dB_\1$.  The seven-dimensional equations are
%%%%%
\bea
R_{\mu\nu} &=& \ft12\del_\mu\phi\, \del_\nu\phi + 
\ft12 e^{2\sqrt{\fft35}\,\phi}\,(G^2_{\mu\nu} 
                        - \ft1{10}\, G_\2^2\, g_{\mu\nu})\,,\nn\\
\square\, \phi &=& \ft12\sqrt{\ft35}\, e^{2\sqrt{\fft35}\,\phi}\, G_\2^2
\,,\label{eimadi}\\
d\Big( e^{2\sqrt{\fft35}\,\phi}\, {*G_\2}\Big) &=& 0\,.\nn
\eea
%%%%%

   Comparing (\ref{8ans}) and (\ref{kkred}), we see that 
%%%%%
\bea
&&B_\1 = \cA_\1\,,\qquad e^{\sqrt{\fft53}\, \phi} = b^2\,,\nn\\
&&ds_7^2 = b^{2/5}\, (dt^2 + a^2\, (D\mu^i)^2 + c^2\, d\Omega_4^2)\,.
\label{seven}
\eea
%%%%%
It is easily verified that the field equation for $G_\2$ given in
(\ref{eimadi}) is automatically satisfied.  The metric $ds_7^2$ lies
within the class whose Ricci tensor was calculated in
\cite{gibpagpop}, and so using those results it is now a straightforward
to obtain the equations for the functions $a$, $b$ and $c$ that follow
from imposing eight-dimensional Ricci-flatness.

   It is convenient to express the equations for $a$, $b$ and $c$ as a
Lagrangian system.  We find that the equations can be derived from varying
$L\equiv T-V$ where
%%%%%
\bea
T &=& 2{\a'}^2 + 12 {\gamma'}^2 + 4\a'\, \beta' + 8 \beta'\, \gamma' +
    16 \a'\, \gamma'\,,\nn\\
V &=& \ft12  b^2\, c^4\, (4a^6 + 2a^4\, b^2 - 24 a^4 c^2 -4 a^2 c^4 + 
    b^2 \, c^4)\,,\label{tveq}
\eea
%%%%%
together with the constraint $T+V=0$.  Here a prime denotes a derivative 
with respect to a new radial variable $\eta$, defined by 
$dt=a^2\, b\, c^4\, d\eta$, and we have also defined $\a=\log a$, 
$\beta = \log b$,  $\gamma=\log c$.  

  We find that the potential $V$ can be derived from a superpotential $W$.  
Writing $T=\ft12 g_{ij}\, (d\a^i/d\eta)\, (d\a^j/d\eta)$, 
where $\a^i=(\a,\beta,\gamma)$,  we have 
$V=-\ft12 g^{ij}\, (\del W/\del\a^i)\, 
(\del W/\del\a^j)$, where
%%%%%
\be
W = b\, c^2\, (4a^3 + 2 a^2\, b + 4 a\, c^2-b\, c^2)\,.
\ee
%%%%%
From this we can obtain the first-order equations $d\a^i/d\eta = g^{ij}\, 
\del W/\del\a^j$.  Expressed back in terms of the original radial variable
$t$ introduced in (\ref{8ans}), these equations are
%%%%%
\be
\dot a = 1 -\fft{b}{2a} - \fft{a^2}{c^2}\,,\qquad
\dot b = \fft{b^2}{2a^2} - \fft{b^2}{c^2}\,,\qquad
\dot c = \fft{a}{c} + \fft{b}{2c}\,,\label{firstorder}
\ee
%%%%%
where a dot denotes a derivative with respect to $t$.

    Before proceeding to find new solutions to these first-order
equations, we can first easily verify that the
previous Spin(7) metric (\ref{spin7metric}) is indeed a solution.
Also, we may observe that one of the seven-dimensional metrics of
$G_2$ holonomy has principal orbits that are $\CP^3$, written as an
$S^2$ bundle over $S^4$, and is given by \cite{brysal,gibpagpop}
%%%%%
\be
ds_7^2 = (1-\fft{\ell^4}{r^4})^{-1}\, dr^2 + \ft14 r^2\, 
(1-\fft{\ell^4}{r^4}) \, (D\mu^i)^2 + \ft12 r^2\, d\Omega_4^2\,.
\ee
%%%%%
This is a solution of the seven-dimensional equations (\ref{eimadi})
with $B_\1=0$ and $\phi=0$, and therefore gives a solution in $D=8$ of
the form $d\hat s_8^2 = ds_7^2 + d\varphi^2$.  This can be described
within the framework of our first-order equations (\ref{firstorder})
by first rescaling $b\longrightarrow\lambda\, b$, and then sending
$\lambda$ to zero, so that the gauge potential $B_\1$ disappears and
$b=$constant is allowed as a solution.\footnote{In appendix A we show
how this ${\cal M}_7\times S^1$ metric arises as a limit of a general 
class of Spin(7) manifolds.}

   One can also see the specialisations to the previous results
described above at the level of the first-order equations themselves.
Setting $a=b$ gives a consistent truncation of (\ref{firstorder}),
yielding $\dot a= \ft12 a^2\, c^{-2}$, $\dot c= \ft32 a\, c^{-1}$,
which are indeed the first-order equations for the original Spin(7)
metrics.  On the other hand, sending $b\longrightarrow 0$ in
(\ref{firstorder}) yields a consistent truncation to $\dot a =1-a^2\,
c^{-2}$, $\dot c= a\, c^{-1}$, which are the first-order equations for
the metrics of $G_2$ holonomy whose principal orbits are $S^2$ bundles
over $S^4$.  (The first-order equations for these two cases can be
found, for example, in \cite{cglpcal}.)

   Another specialisation of the metric ansatz (\ref{8ans}) that makes
contact with previous results is to set $a=c$, in which case the $S^2$
bundle over $S^4$ becomes precisely the usual $\CP^3$ Einstein
manifold, with its $SU(4)$-invariant metric.  This is incompatible
with the first-order equations (\ref{firstorder}), but it is easily
verified that it is consistent with the second-order Einstein
equations following from (\ref{tveq}).  Solutions to these
second-order equations then include the 8-dimensional Taub-NUT and
Taub-BOLT metrics.  The incompatibility with the first-order equations
is understandable, since the Taub-NUT and Taub-BOLT 8-metrics do not
have special holonomy.  Another previously-seen solution of the
second-order equations with $a=c$ is the Ricci-flat K\"ahler metric on
the complex line-bundle over $\CP^3$.  Although this can arise from a
first-order system, it is an inequivalent one that is not related to a
specialisation of (\ref{firstorder}).  Its superpotential is $W=2 a^6
+ 6 a^4\, b^2$ \cite{cglp1}, with $T$, $V$ and $g_{ij}$ following from
setting $a=c$ in (\ref{tveq}).  (Other examples of this kind of
phenomenon were exhibited recently in \cite{cglpcal}.)

\section{Solving the Ricci-flat equations}

    In order to obtain new solutions of the first-order equations 
(\ref{firstorder}) we first introduce a new radial 
coordinate $r$, defined in terms of $t$ by $dr = b\, dt$.  After also defining
$f\equiv c^2$, we find by taking further derivatives of the first-order
equations (\ref{firstorder}) that $f$ must satisfy the third-order 
equation
%%%%%
\be
2 f^2\, f''' + 2 f\,(f'-3)\, f'' - (f'+1)(f'-1)(f'-3)=0\,.
\label{feq}
\ee
%%%%%
The remaining metric functions are then given by solving
%%%%%
\be
a'=\fft{f'-2}{2a} -\fft{(f'-1)\, a}{2f}\,,\qquad
b^2 = \fft{4 a^2}{(f'-1)^2}\,.\label{absol}
\ee
%%%%%
Naively there now appear to be four constants of integration in total rather
than the expected three, but the extra one is eliminated by 
substituting the solutions back into (\ref{firstorder}).  

   We have found two simple independent non-trivial
solutions\footnote{The general solution is constructed in appendix A.
It gives further inequivalent regular metrics, complete on 
manifolds $\bB_8^\pm$.  These solutions are more complicated, but still fully 
explicit (up to quadratures).} to (\ref{feq}), 
which can be reduced to $f=3r$ and $f=r+
r^2/(2\ell^2)$.  The solution with $f=3r$ implies $a^2=b^2=\ft35 r +
k\, r^{-2/3}$, and after performing the coordinate transformation
$r\longrightarrow 3 r^2/20$ this gives precisely the previously-known
Spin(7) solution (\ref{spin7metric}), with $\ell=k^{3/10}\,
(20/3)^{1/5}$.

  Our new simple solutions of Spin(7) holonomy arise from the second
solution, $f=r+ r^2/(2\ell^2)$.  After making the coordinate
transformation $r\longrightarrow -\ell\, (r+\ell)$, this solution
leads to the metric
%%%%%
\be
ds_8^2 = \fft{(r+\ell)^2\, dr^2}{(r+3\ell)(r-\ell)} +
\fft{\ell^2\, (r+3\ell)(r-\ell)}{(r+\ell)^2}\, \sigma^2 
+ \ft14(r+3\ell)(r-\ell)\, (D\mu^i)^2 + \ft12(r^2-\ell^2)\, d\Omega_4^2\,,
\label{sol1}
\ee
%%%%%

    Assuming that the constant $\ell$ is positive, it is evident that
$r$ should lie in the range $r\ge\ell$.  We can analyse the behaviour
near $r=\ell$ by 
defining a new radial coordinate $\rho$, where 
$\rho^2=4\ell\,(r-\ell)$.  Near $\rho=0$ the metric approaches
%%%%%
\be
ds_8^2 \approx d\rho^2 + \ft14 \rho^2 \, \Big[\sigma^2 + (D\mu^i)^2 + 
d\Omega_4^2\Big]\,.
\ee
%%%%%
The quantity $\ft14(\sigma^2+(D\mu^i)^2 + d\Omega_4^2)$  
is precisely the metric on the unit
7-sphere, and so we see that near $r=\ell$ the metric $ds_8^2$
smoothly approaches flat $\R^8$.  At large $r$ the function $b$, which
is the radius in the $U(1)$ direction $\sigma$, approaches a constant,
and so the metric approaches an $S^1$ bundle over a 7-metric.  This
7-metric is of the form of a cone over ${\Bbb C} {\Bbb P}^3$
(described as the $S^2$ bundle over $S^4$) in this asymptotic
region.  The manifold of this new Spin(7) metric, which we are
denoting by $\bA_8$,  is topologically $\R^8$.

   We shall use the acronym AC to denote asymptotically conical
manifolds. Thus asymptotically our new metrics behave like a circle
bundle over an AC manifold in which the length of the $U(1)$ fibres
tends to a constant. The acronym ALF is already in use to describe
metrics which tend to a $U(1)$ bundle over an asymptotically Euclidean
or asymptotically locally Euclidean metric with the length of the
fibres tending to a constant. We shall therefore adopt the acronym ALC
to denote manifolds where the base space of the circle bundle is
asymptotically conical. 

   Ricci-flat ALC metrics, although not with special holonomy, have already
been encountered.  For example, the higher-dimensional Taub-NUT metric
is defined on $\R^{2n}$ for all $n$ and it is ALC with the base of the
cone being $\CP^{n-1}$. A closely related example is
the Taub-BOLT metric which has the same asymptotics but is defined
on a line bundle over $\CP^{n-1}$. However, as we shall
see later, the metric on the base of the cone differs in this case
(with $n=4$) from that in our new metrics.  An discussion of
ALE Spin(7) manifolds based on the idea of blowing up orbifolds has
been given in \cite{joyce}.  As far as we are aware, no explicit
examples of this kind have yet been found.

   We get a different complete manifold, which we are denoting by
$\bB_8$, if we take $r$ to be negative.
It is easier to discuss this by instead setting $\ell=-\td\ell$, where
$\td\ell$ and $r$ are taken to be positive.  
Thus instead of (\ref{sol1}) we now have
%%%%%
\be
ds_8^2 = \fft{(r-\td\ell)^2\, dr^2}{(r-3\td\ell)(r+\td\ell)} +
\fft{\td\ell^2\, (r-3\td\ell)(r+\td\ell)}{(r-\td\ell)^2}\, \sigma^2 
+ \ft14(r-3\td\ell)(r+\td\ell)\, (D\mu^i)^2 + 
\ft12(r^2-\td\ell^2)\, d\Omega_4^2\,,
\label{sol2}
\ee
%%%%%
This time, we have $r\ge 3\td\ell$.  
Defining $\rho^2= 4\td\ell\, (r-3\td\ell)$, 
we find that near $r=3\td\ell$ the metric has the form
%%%%%
\be
ds_8^2 \approx d\rho^2 + \ft14 \rho^2\, [\sigma^2 + (D\mu^i)^2] + 4\td\ell^2\, 
d\Omega_4^2\,.
\ee
%%%%%
The quantity $\ft14[\sigma^2 + (D\mu^i)^2]$ is the metric on the unit
3-sphere, and so in this case we find that the metric smoothly
approaches $\R^4\times S^4$ locally, at small distance.  The
large-distance behaviour is the same as for the previous case
(\ref{sol1}).

   Again we have a complete non-compact ALC metric with Spin(7)
holonomy with the same base.  At short distance, it has the same
structure as the previously-known metric of Spin(7) holonomy, obtained
in \cite{gibpagpop}. Thus globally the manifold $\bB_8$ is the bundle of
chiral spinors over $S^4$.

   We can think of the new manifold $\bA_8$ as providing a smooth
intepolation between Euclidean 8-space at short distance, and 
${\cal M}_7\times S^1$ at large distance, while $\bB_8$ provides an
interpolation between the previous Spin(7) manifold of
\cite{brysal,gibpagpop} at short distance and ${\cal M}_7\times S^1$
at large distance.   Here ${\cal M}_7$ denotes the 7-manifold of 
$G_2$ holonomy that is the $\R^3$ bundle over $S^4$
\cite{brysal,gibpagpop}.

   In appendix A we construct the general solution of the first-order
equations (\ref{firstorder}).  From this, we find additional classes
of regular metrics of Spin(7), which are complete on manifolds
$\bB_8^\pm$ that are similar to $\bB_8$.  These additional metrics
have a non-trivial integration constant $k$ that parameterises
inequivalent solutions.

   It is worth remarking that we would obtain identical equations to
solve if we were to replace the $S^4$ metric $d\Omega_4^2$ in
(\ref{8ans}) by the Fubini-Study metric on $\CP^2$, scaled so that it
has the same cosmological constant as the unit 4-sphere.  (In fact the
first-order equations in this case are contained within those obtained
in \cite{cglpcal}.)  The Yang-Mills connection $A_\1^i$ would now be
the right-handed projection of the spin connection on $\CP^2$.
However, the analogue of the $\bA_8$ manifold would now have power-law
singularities in the Riemann tensor at $r=\ell$, since the principal
orbits that collapse to a point would be $SU(3)/U(1)$ instead of
$S^7$.  The analogue of the $\bB_8$ manifold would not have power-law
curvature singularities at $r=3\td\ell$, but it would have an orbifold
singularity there, approaching $(\R^4/\Z_2)\times \CP^2$ locally.  The
reason for this is that the Yang-Mills connection on $\CP^2$ is in
$SO(3)$ rather than $SU(2)$, and so the collapsing 3-surfaces at
$r=3\td\ell$ will be $\RP^3$ rather than $S^3$.

\section{Proof of Spin(7) holonomy}

   Our procedure for solving the condition of Ricci-flatness for the
eight-dimensional metric ansatz (\ref{8ans}) involved establishing that
there exists a superpotential for the potential in the Lagrangian
formulation of the Einstein equations, and hence obtaining the first-order
equations (\ref{firstorder}).  The fact that such a first-order system
exists provides a strong indication that there is an underlying special
holonomy, since such systems of equations typically arise from the 
conditions for the covariant constancy of a spinor.  However, it is
still necessary to make a more thorough investigation in order to 
establish definitively that our new solutions have Spin(7) holonomy.

  A convenient way to study this question is by again making use of
the Kaluza-Klein reduction (\ref{kkred}), so that the equation $\hat
D\, \eta=0$ for a covariantly-constant spinor in $D=8$ can be
reformulated in $D=7$.  (Here $\hat D\equiv d+ \ft14
\hat\omega_{AB}\, \Gamma_{AB}$ is the Lorentz-covariant exterior
derivative that acts on spinors in eight dimensions, where $\Gamma_{AB}
\equiv \ft12(\Gamma_A\, \Gamma_B-\Gamma_B\, \Gamma_A)$, and 
$\Gamma_A$ are the Dirac matrices that generate the 
Clifford algebra in eight dimensions.)  The advantage
of doing this is that we can then make use of results derived in
\cite{gibpagpop} for the spin connection for 7-metrics of the type
given in (\ref{seven}).  Specifically, we find that under Kaluza-Klein
reduction we have
%%%%%
\bea
\hat D&=& D +\ft1{4\sqrt{15}}\, \del_a\phi\, \Gamma_{ab}\, e^b 
-\ft18 \cF_{ab}\, e^{2\sqrt{\fft35}\, \phi}\, \Gamma_{ab}\,
(d\varphi+\cA_\1) \nn\\
&&- \ft14\sqrt{\ft53}\, e^{\sqrt{\fft35}\, \phi}\,  
\del_a\phi\, \Gamma_{a8} \, (d\varphi+\cA_\1) 
-\ft14 e^{2\sqrt{\fft35}\, \phi}\, \Gamma_{a8}\, e^b\,,
\eea
%%%%%
where $D\equiv d+\ft14 \omega_{ab}\, \Gamma_{ab}$ is the Lorentz-covariant
exterior derivative that acts on spinors in seven dimensions, and
$\omega_{ab}$ can be read off from \cite{gibpagpop}.  

   Using the results in \cite{gibpagpop} for the spin connection for 
7-metrics of the form appearing in (\ref{seven}), we eventually find that
if and only if the metric functions $a$, $b$ and $c$ satisfy the
the first-order equations (\ref{firstorder}), then   
the eight-dimensional equation $\hat D\, \eta=0$ has exactly one solution.
The solution for the covariantly-constant spinor $\eta$ can be written as
%%%%%
\be
\eta = e^{\fft12 \theta\, \Gamma_{71}}\, e^{\fft12\psi\, \Gamma_{12}}\, 
\eta_0\,,
\ee
%%%%%
where $\eta_0$ is independent of $(r,\theta,\psi,\varphi)$, and satisfies
projection conditions that are all implied by
%%%%%
\be
(\Gamma_{12}-\Gamma_{78})\,\eta_0=0\,,\quad 
(F^3_{\a\beta}\, \Gamma_{\a\beta} +4\Gamma_{78})\, \eta_0=0\,,\quad
(F^1_{\a\beta}\, \Gamma_{\a\beta}+ 4\Gamma_{71})\, \eta_0 =0\,.
\ee
%%%%%
Here the tangent-space indices 1 and 2 lie in the $S^2$ directions, 
$(\a,\beta)$ lie in the $S^4$ directions, 7 is in the radial direction, and
8 is in the $U(1)$ fibre direction.  In our conventions, the Yang-Mills 
instanton fields $F_\2^i$ on $S^4$ are given by
%%%%%
\be
F_\2^1 = -(e^4\wedge e^5 + e^3\wedge e^6)\,,\quad
F_\2^2 = -(e^5\wedge e^3 + e^4\wedge e^6)\,,\quad
F_\2^3 = -(e^3\wedge e^4 + e^5\wedge e^6)\,,
\ee
%%%%%
where $e^\a=(e^3,e^4,e^5,e^6)$ is the basis of tangent-space 1-forms on the 
unit $S^4$.  The spinor $\eta_0$ satisfies the equations for the zero-mode of 
the Dirac equation on $S^4$ in the Yang-Mills instanton background.

    With these results, we have established that the first-order equations
(\ref{firstorder}) are indeed the integrability conditions for the
existence of a single covariantly-constant spinor in the 8-metric 
(\ref{8ans}).  This establishes that for any solution of (\ref{firstorder}), 
we obtain an 8-metric (\ref{8ans}) that has Spin(7) holonomy.  The
existence of the covariantly-constant spinor $\eta$ immediately implies
the existence of a covariantly-constant self-dual 4-form $\Phi$, with
components given by $\Phi_{ABCD} = \bar\eta\, \Gamma_{ABCD}\, \eta$.
The covariant constancy of $\eta$ implies that $\bar\eta\, \eta$ is
constant, and so we may choose a normalisation so that $\bar\eta\,
\eta = 1$.  We then find that the 4-form is given by
%%%%%
\bea
\Phi &=& -\hat e^1\wedge \hat e^2\wedge \hat e^7\wedge \hat e^8 - 
   \hat e^3\wedge \hat e^4\wedge \hat e^5\wedge \hat e^6 
+ (\hat e^1\wedge \hat e^2+\hat e^7\wedge \hat e^8) 
\wedge \hat Y_\2 \nn\\
&&+ (\hat e^1\wedge \hat e^8+ \hat e^2\wedge \hat e^7)\wedge 
\fft{\del \hat Y_\2}{\del \theta}
 -(\hat e^1\wedge \hat e^7 - \hat e^2\wedge \hat e^8)\wedge 
\fft{1}{\sin\theta}\, 
\fft{\del \hat Y_\2}{\del\psi}\,,\label{phiexp}
\eea
%%%%%
where $\hat Y_\2\equiv c^2\, Y_\2$ and $Y_\2\equiv \mu^i\, F_\2^i$, 
and so 
%%%%%
\be
\hat Y_\2 = \ft12[\sin\theta\, (\cos\psi\, F_{\a\beta}^1 + 
\sin\psi\, F_{\a\beta}^2)
        + \cos\theta\, F_{\a\beta}^3]\, \hat e^\a\wedge \hat e^\beta\,,
\ee
%%%%%
where as usual $\hat e^\a=c\, e^\a$.  

    The covariantly-constant self-dual 4-form $\Phi$, known as the
Cayley form, provides a calibration of the Spin(7) manifold.  Thus we have
%%%%%
\be
|\Phi(X_1,X_2,X_3,X_4)| \le 1\,,
\ee
%%%%%
where $(X_1,X_2,X_3,X_4)$ denotes any quadruple of orthonormal
vectors.  This can be seen from (\ref{phiexp}), or else from the
expression $\Phi_{ABCD} = \bar\eta\, \Gamma_{ABCD}\, \eta$.  A
calibrated submanifold, or Cayley submanifold, $\Sigma$, is one where
for each point of $\Sigma$
%%%%%
\be
|\Phi(X_1,X_2,X_3,X_4)|=1\,,
\ee
%%%%%
where the orthonormal vectors $X_i$ are everywhere tangent to
$\Sigma$.  By inspecting (\ref{phiexp}) we therefore see that the
$S^4$ zero section of the bundle of chiral spinors is a Cayley
submanifold, and hence it is volume minimising in its homology class.
Physically, a Cayley submanifold corresponds to a supersymmetric cycle 
\cite{bbmooy,gibpap}.

\section{$L^2$-normalisable harmonic 4-forms}

     In this section, we obtain $L^2$ normalisable harmonic 4-forms
for each of the new Spin(7) 8-manifolds $\bA_8$ and $\bB_8$.
Specifically, we obtain one such 4-form, which is anti-self-dual, for the
manifold $\bA_8$ that is topologically $\R^8$, and two such 4-forms, one
of each duality, for the manifold $\bB_8$ of the chiral spin bundle over
$S^4$.

   We start from the following ansatz for the harmonic 4-forms,
%%%%%%
\bea
G_\4 &=& u_1\, (h\, a^2\, b\, dr\wedge \sigma\wedge X_\2 \pm c^4\,
\Omega_\4) + u_2\, (h\, b\, c^2\, dr\wedge \sigma\wedge Y_\2 \pm a^2\,
c^2\, X_\2\wedge Y_\2)\nn\\
&&+ u_3\, (h\, a\, c^2\, dr\wedge Y_\3 \mp 
b\, a\, c^2\,\sigma\wedge X_\3)\,,\label{4formans}
\eea
%%%%%%
where $\Omega_\4$ is the volume form of the unit $S^4$, and 
%%%%%
\bea
&&X_\2\equiv \ft12\epsilon_{ijk}\, \mu^i\, D\mu^j\wedge D\mu^k\,,\qquad
X_\3\equiv D\mu^i\wedge F_\2^i\,,\nn\\
&&Y_\2\equiv\mu^i\,F_\2^i\,,\qquad
Y_\3\equiv\epsilon_{ijk}\, \mu^i\, D\mu^j\wedge F_\2^k\,.\label{23forms}
\eea
%%%%%
The upper and lower sign choices in (\ref{4formans}) correspond to
self-dual and anti-self-dual 4-forms respectively. 
The various 2-forms and 3-forms defined in (\ref{23forms}) satisfy
%%%%
\be
d\sigma=X_\2 - Y_\2\,,\quad dX_\2=X_\3=dY_\2\,,\qquad
dY_\3=2 X_\2\wedge Y_\2 + 4\Omega_\4\,.
\ee
%%%%
Note that in (\ref{4formans}) we have introduced a radial coordinate
$r$ that is related to $t$ by $dt=h\, dr$.

    $G_\4$ will be harmonic if $dG_\4=0$.  This implies that
%%%%%
\bea
(c^4\, u_1)' &=& \pm 2(-h\, b\, c^2\, u_2 + 2 h\, a\, c^2\,
u_3)\,,\nn\\
(a^2\, c^2\, u_2)' &=& \pm(- h\, a^2\, b\, u_1 +
h\, b\, c^2\,  u_2  + 2h\, a\, c^2\, u_3)\,,\label{4formeq}\\
(a\, b\, c^2\, u_3)' &=& \pm (h\, a^2\, b\,  u_1 + h\, b\, c^2\, u_2)
\,.\nn
\eea
%%%%%%
The $\pm$ signs correspond to
self-dual and anti-self-dual respectively, and a prime denotes a
derivative with respect to $r$.  In the remainder of this section, we
shall for convenience set the scaler parameters $\ell$ and $\td\ell$
in the metrics (\ref{sol1}) and (\ref{sol2}) to unity.\footnote{Care
must be exercised when taking the square roots of $a^2$, $b^2$ and
$c^2$ in the metrics (\ref{sol1}) and (\ref{sol2}), if one wants the 
functions $a$, $b$ and $c$ to solve precisely the first-order
equations (\ref{firstorder}), since these equations are sensitive to
the signs of $a$, $b$ and $c$.  (Of course there are equivalent
first-order equations that differ by precisely these sign factors, and
which also imply solutions of the Einstein equations.)  We are assuming
here that the signs are chosen so that precisely (\ref{firstorder})
are satisfied.  This can be achieved by taking all square roots to be
positive, except for $b$ in the case of (\ref{sol1}) on $\bA_8$.}

     For the metric (\ref{sol1}) on the manifold $\bA_8$ that is
topologically $\R^8$, we find
that there is a normalisable harmonic 4-form that is anti-self-dual,
{\it i.e.}, the lower choice of the sign is used in (\ref{4formans})
and (\ref{4formeq}).  The solution is given by
%%%%%%%
\be
u_1 =\fft{2}{(r+1)^3(r+3)}\,,\quad
u_2 =-\fft{r^2 +10r + 13}{(r+1)^3(r+3)^3}\,,\quad
u_3 =-\fft2{(r+1)^2(r+3)^3}\,.\label{aasd}
\ee
%%%%%%
The norm of the harmonic anti-self-dual 4-form is then given by
%%%%%
\bea 
|G_\4|^2 &=& 48(u_1^2 +2 u_2^2 + 4 u_3^2)\nn\\ 
&=& \fft{96(3r^4+44r^3
+ 242r^2 + 492r +339)}{(r+1)^6(r+3)^6}\,.\label{g4square1anti}
\eea 
%%%%%
Clearly $G_\4$ is $L^2$-normalisable, and in fact we have
$\int_1^\infty \sqrt{g}\, |G_\4|^2\, dr = 9/4$.  We have chosen
the integration constants from (\ref{4formeq}) appropriately in order
to select the solution in $L^2$.  (There also exists a solution for 
a self-dual harmonic 4-form.  It can be made square integrable at small
distance, but there is no choice of integration constants for which it
is $L^2$ normalisable, owing to its large distance behaviour.)

      For the metric (\ref{sol2}) on $\bB_8$, the bundle of chiral spinors
over $S^4$, we find that there exists a normalisable harmonic 4-form
that is anti-self-dual, {\it i.e.}, the lower choice of sign is used
in (\ref{4formans}) and (\ref{4formeq}).  The solution is given by
%%%%%
\bea
&&u_1=\fft{2(r^4+8r^3+34r^2-48r+21)}{(r-1)^3(r+1)^5}\,\qquad
u_2= - \fft{r^4+4r^3-18r^2+52r-23}{(r-1)^3(r+1)^5}\,,\nn\\
&&u_3=\fft{2(r^2+14r-11)}{(r-1)^2(r+1)^5}\,.\label{basd}
\eea
%%%%%
The square of the anti-self-dual 4-form is given by
%%%%%
\be
|G_\4|^2 = \fft{96(3r^8 +40r^7 + 252r^6 +1064r^5+2506r^4
-12936r^3 +18284r^2 -10824r +2379)}{(r-1)^6(r+1)^{10}}\,,
\label{g4square2anti}
\ee
%%%%%
and its $L^2$-normalisability can be seen by noting that
$\int_3^\infty  \sqrt{g}\, |G_\4|^2\, dr = 189/16$.

     Both of the above harmonic anti-self-dual 4-forms (\ref{aasd})
and (\ref{basd}) on $\bA_8$ and
$\bB_8$ satisfy the linear relation
%%%%%
\be
u_1+2u_2-4u_3=0\,.\label{u123linear}
\ee
%%%%
This observation will prove useful later, for showing the supersymmetry
of resolved brane solutions.

   We also find a second $L^2$-normalisable harmonic 4-form in the
new Spin(7) manifold $\bB_8$. This 4-form is self-dual, and is given by 
%%%%%
\bea
&&u_1= -\fft{2(5r^3-9r^2+15r-3)}{(r-1)^3\, (r+1)^4}\,,\nn\\
&&u_2 = \fft{(r-3)(5r^2-2r+1)}{(r-1)^3\, (r+1)^4}\,,\quad
u_3=-\fft{2(r-3)}{(r-1)^2\, (r+1)^4}\,.\label{bsd}
\eea
%%%%%
In contrast to the previous harmonic 4-forms, there is no linear relation 
between the functions $u_1$, $u_2$ and $u_3$ here.  The magnitude of $G_\4$
is given by
%%%%%
\be
|G_\4|^2 = \fft{96(75r^6-350r^5+829 r^4-932 r^3+885 r^2-414 r+99)}{
               (r-1)^6\, (r+1)^8}\,.\label{g4square2self}
\ee
%%%%%
It integrates to give $\int_3^\infty \sqrt{g}\, |G_\4|^2\, dr =189/4$.

   It is interesting to note that for the anti-self-dual harmonic
4-form on $\bA_8$, given by (\ref{aasd}), we can write it in terms of
a globally-defined potential, $G_\4=dB_\3$.  Specifically, we find
that $B_\3$ can be written as 
%%%%%
\be
B_\3= (r-1)^2\, \Big[ -\fft1{8(r+1)^2}\, \sigma\wedge X_\2 +
\fft{(r+5)}{8(r+1)(r+3)^2}\, \sigma\wedge Y_\2 -\fft{1}{16 (r+3)^2}\,
   Y_\3\Big]\,.\label{b3exp}
\ee
%%%%%
One can see from (\ref{sol1}) that this has a vanishing magnitude
$|B_\3|^2$ at $r=1$.  On the other hand the analogous expressions for
the potential $B_\3$ for the two harmonic 4-forms (\ref{basd}) and
(\ref{bsd}), which are similarly expressible as functions of $r$ times
the three 3-form structures in (\ref{b3exp}), turn out to have a
diverging magnitude at $r=3$.  In all three cases the $r$-dependent
prefactors tend to constants at infinity.

\section{Comparison with Taub-NUT and Taub-BOLT metrics}

   As mentioned above, the new 8-metrics of Spin(7) holonomy that we
have obtained in this paper have an asymptotic large-distance
behaviour that is similar to the one seen in the 8-dimensional
Taub-NUT and Taub-BOLT metrics.  Unlike those metrics, however, ours
admit a covariantly-constant spinor, and so they have special holonomy
Spin(7).

    It is worthwhile  looking at the 
comparison with the Taub-NUT and Taub-BOLT 8-metrics in a little
more detail.  The 8-dimensional Taub-NUT metric can be written as
(see, for example, \cite{baisbate,awacha,cglpcal})
%%%%%
\be
ds_8^2  = \fft{5(r+\ell)^3}{8(r-\ell)(r^2+4\ell\, r+5\ell^2)}\, dr^2
  + \fft{8(r-\ell)(r^2+4\ell\, r+5\ell^2 )}{5 (r+\ell)^3}\,
\sigma^2 + (r^2-\ell^2)\, d\Sigma_3^2\,,\label{d8tn}
\ee
%%%%%
where $d\Sigma_3^2$ is the Fubini-Study metric on the ``unit'' $\CP^3$,
and $\sigma=d\varphi+A_\1$, where $dA_\1=2J$ and $J$ is the K\"ahler form on 
$\CP^3$.  The Taub-BOLT metric can be written as \cite{awacha}
%%%%%
\be
ds_8^2  = { dr^2 \over F(r)}
  + F(r)\,
\sigma^2 + (r^2-\ell^2)\, d\Sigma_3^2\,,\label{d8tb}
\ee
%%%%%
where 
%%%%
\be
F= {8( r^6 -5\ell^2\, r^4 + 15 \ell^4 r^2 -10 m\,r + 
5\ell^6) \over 5(r^2 -\ell^2 )^3 }\,,
\ee
%%%%
and we choose the integration constant $m=\ft85\ell^5$.  This choice
means that $F$ vanishes at $r=4\ell$, which is a smooth 6-dimensional
fixed point set (geometrically a $\CP^3$) of the $U(1)$ action.

   The general Taub-BOLT and Taub-NUT metrics are thus constructed as
metrics of cohomogeneity one with principal orbits that are $\CP^3$
with its standard $SU(4)$-invariant Fubini-Study metric.  By contrast,
although our new metrics are again of cohomogeneity one with $\CP^3$
principal orbits, the metric on $\CP^3$ is ``squashed,'' and is
constructed as an $S^2$ bundle over $S^4$, with isometry group
$SO(3)\times SO(5)$.  At large distance our new solutions are ALC,
\ie of the form of an $S^1$ bundle over the cone with base the
``squashed'' Einstein metric on $\CP^3$.  By contrast, the $D=8$
Taub-NUT and Taub-BOLT metrics are asymptotically cylindrical, having
the form of an $S^1$ bundle over the cone with base the ``round''
Fubini-Study metric on $\CP^3$.

   At short distance, our solution approaches either $\R^8$ or
$\R^4\times S^4$ locally, depending on whether the parameter $\ell$ in
(\ref{sol1}) is taken to be positive or negative.  The manifold
$\bA_8$ that approaches $\R^8$ is very similar in its short-distance
behaviour to the 8-dimensional Taub-NUT metric, and indeed both
metrics are defined on $\R^8$. On the other hand, the 8-dimensional
Taub-BOLT locally approaches $\R^2\times \CP^3$ at short distance.
Thus while our solution on $\bB_8$, given in (\ref{sol2}), which
approaches $\R^4\times S^4$ locally at short distance, could be
thought of as somewhat analogous to $D=8$ Taub-BOLT, it is of a quite
different structure. In four dimensions the terms NUT and BOLT were
originally defined \cite{GH} as zero-dimensional and two-dimensional
fixed point sets of a $U(1)$ action. They are not infrequently
extended to cover the more general case of the degenerate orbits of a
higher-dimensional isometry group $G$, say. These are the orbits which
are smaller in dimension than the generic or principal orbits. However
a subtlety now arises because (even in four dimensions) such
degenerate orbits may or may not be the fixed point sets of a $U(1)$
subgroup of the isometry group $G$. In the present cases the
degenerate orbit is also the fixed point set of a circle action, and
so the original and the extended meaning both apply. In all the cases
we consider, the circle action is generated by the Killing field
$\partial \over \partial \varphi$, and since its length squared is
$g({\partial \over \partial \varphi} ,{\partial \over \partial
\varphi} )=b^2 =F$, it has a fixed point set when $b$ or $F$ vanishes.
   
   A further point of interest concerns the feature of our new 
Spin(7) metric obtained in section 3 
that it can be defined on two inequivalent regular
non-compact manifolds, depending on whether $r$ is positive or negative.  
(We equivalently presented the choice in terms of an $r$ that is always 
positive, but with opposite signs for the scale parameter $\ell$.)  In fact 
although this feature is somewhat unusual, it does also occur in at least one
other previously-known metric.  Specifically, the 6-dimensional 
Taub-NUT metric can be written as 
%%%%%
\be
ds_6^2  = \fft{(r+\ell)^2}{2(r-\ell)(r+3\ell)}\, dr^2 +
 \fft{2(r-\ell)(r+3\ell)}{(r+\ell)^2}\,
\sigma^2 + (r^2-\ell^2)\, d\Sigma_2^2\,,\label{tn1}
\ee
%%%%%
where $d\Sigma_2^2$ is the Fubini-Study metric on the unit $CP^2$, 
and $\sigma=d\varphi+A$ with $dA=2J$ (see, for example, 
\cite{baisbate,awacha,cglpcal}).  If one takes $\ell$ to be positive,
then this metric is defined for $r\ge\ell$, and near $r=\ell$ it
approaches $\R^6$.  This can be seen by letting $\rho^2=2\ell\, (r-\ell)$, 
so that near $\rho=0$ we have $ds_6^2\approx d\rho^2 + \rho^2\, 
(\sigma^2+d\Sigma_2^2)$, and $\sigma^2 + d\Sigma_2^2$ can be recognised
as the metric on the unit $S^5$ described as a $U(1)$ bundle over $\CP^2$,
provided that $\varphi$ has period $4\pi$. 

   If, on the other hand, we set $\ell= -\td\ell$, so 
that the metric becomes 
%%%%%
\be
ds_6^2  = \fft{(r-\td\ell)^2}{2(r+\td\ell)(r-3\td\ell)}\, dr^2 +
 \fft{2\td\ell^2\, (r+\td\ell)(r-3\td\ell)}{(r-\td\ell)^2}\,
\sigma^2 + (r^2-\td\ell^2)\, d\Sigma_2^2\,,\label{tn2}
\ee
%%%%%
then we now have $r\ge 3\td\ell$.  Near to $r=3\td\ell$ we can 
introduce a new radial coordinate such that $\rho^2=2\td\ell(\, r-3\td\ell)$, 
and so  the metric approaches
%%%%%
\be
d\hat s_6^2\approx d\rho^2 + \rho^2\, \sigma^2 + 8 \td\ell^2\, d\Sigma_2^2\,.
\ee
%%%%%
Regularity at $\rho=0$ requires that the coordinate $\varphi$ should
have period $\Delta \varphi=2\pi$.  The period that would be needed
for the $U(1)$ bundle over $CP^2$ to be $S^5$ is $\Delta \varphi =
4\pi$.  Therefore the level surfaces of the principal orbits are
$S^5/\Z_2$.  The metric smoothly approaches $\R^2\times \CP^2$ locally
at short distance, and has the usual cylindrical Taub-NUT form at
large $r$.  Somewhat surprisingly, we find that this is in fact
precisely the $D=6$ Taub-BOLT solution.  Thus we have the remarkable
result that in $D=6$ the Taub-BOLT metric is nothing but the Taub-NUT
metric, seen from the other side of $r=0$.

   This feature of the 6-dimensional Taub-NUT solution, of admitting a
different global interpretation for the opposite sign of $r$ or the scale
parameter $\ell$, does not appear to extend to the Taub-NUT metrics in
$D\ge8$.  For example, in (\ref{d8tn}) there is no additional real
root of the radial function multiplying $\sigma^2$, analogous to the
$(r+3\ell)$ factor in (\ref{tn1}).  Although a Taub-BOLT solution
exists in $D=8$, it is given by the quite different metric
(\ref{d8tb}).  In $D=10$ there does exist another real root, but it
corresponds to an un-removable conical singularity since it would
require that the fibre coordinate $\varphi$ have an irrational period.
As with all the higher-dimensional cases there does also exist a
Taub-BOLT solution, but it has a quite different form.  It seems
likely that the feature of a Taub-NUT metric having a second regular
manifold corresponding to the negative-$r$ region is peculiar to the
six-dimensional case, and it happens neither in $D=4$ nor in $D\ge 8$.
Only for $D=6$ is Taub-NUT its own Taub-BOLT.

\section{Applications in M-theory and string theory}

   The new Spin(7) manifolds have a variety of
applications in M-theory and string theory.  For the present purposes,
these can be discussed as the level of the classical low-energy
effective supergravity field theories.  The bosonic Lagrangian for the
$D=11$ supergravity limit of M-theory is
%%%%%
\be
{\cal L}_{11} = R\, {*\oneone} - \ft12 {*F_\4}\wedge F_\4 +\ft16 
F_\4\wedge F_\4\wedge A_\3\,,\label{d11lag}
\ee
%%%%%
where $F_\4=dA_\3$.  The low-energy limit for type IIA string theory
follows by performing a Kaluza-Klein dimensional reduction of
(\ref{d11lag}) on a circle.

\subsection{D6-branes as Spin(7) manifolds}

   The new Spin(7) manifolds $\bA_8$, $\bB_8$ and $\bB_8^\pm$ provide new
supersymmetric vacua in $D=11$ M-theory, simply by taking the direct
product with a three-dimensional Minkowski spacetime $M_3$, and
setting $F_\4=0$.  We can then dimensionally reduce the solution on
the $\varphi$ fibre coordinate, using the $11\longrightarrow 10$
analogue of (\ref{kkred}), to give a wrapped
D6-brane in type IIA string theory:
%%%%%%%
\bea
ds_{\rm str}^2 &=& \fft{b}{N}\Big(-dt^2 + dx_1^2 + dx_2^2 +
c^2\, d\Omega_4^2 + h^2\, dr^2 + a^2\, D\mu^i\, D\mu^i\Big)\,,\nn\\
e^{\ft43 \phi}&=&\fft{b^2}{N^2}\,,\qquad
\cF_\2 = N\, \Big(\ft1{2} \ep_{ijk}\, \mu^k\, D\mu^i \wedge
D\mu^j - \mu^i\, F_\2^{i}\Big) \,.\label{2ff}
\eea
%%%%%%
Here $ds_{\rm str}^2$ is the string-frame metric in $D=10$, related to
the Einstein-frame metric $ds_{10}^2$ by $ds_{\rm str}^2 =
e^{\fft12\phi}\, ds_{10}^2$.  The string coupling constant is given by
$g=e^{\phi_0}$, where $\phi_0$ is the asymptotic value of $\phi$ at
large distance; $g=(\ell/N)^{3/2}$.  We have introduced an integer $N$
which is the number of D6-branes.  This corresponds to the $D=11$
solution with the $\varphi$ fibre coordinate having a period of
$4\pi/N$.  (There will be an orbifold singularity at the origin if
$N\ne1$.)  The solution can be viewed as D6-branes wrapped around the
4-sphere.  At small distance, the wrapping 4-sphere either collapses
or stablises to a fixed radius, depending on which of our two
manifolds is used.  Using the $\bA_8$ manifold we have an
interpolation from $M_{11}$ at short distance to $M_3\times S^1\times
{\cal M_7}$ at large distance, while for $\bB_8$ or $\bB_8^\pm$ the
interpolation is from $M_3\times {\cal M}_8$ at short distance to
$M_3\times S^1\times {\cal M}_7$ at large distance.  Here $M_n$
denotes $n$-dimensional Minkowski spaectime, ${\cal M}_7$ is the
manifold of $G_2$ holonomy on the $\R^3$ bundle overe $S^4$, and
${\cal M}_8$ is the previously-known manifold of Spin(7) holonomy on
the $\R^4$ bundle over $S^4$.  The world volume at large distance
becomes
%%%%%%
\be
M_3\times S^1\,,
\ee
%%%%%
with the string coupling constant being $g_{\rm str}=R^{3/2}$, where
$R$ is the radius of $S^1$.  Taking $g_{\rm str}$ large implies a
decompactification of $S^1$, thus rendering the world-volume theory to
be effectively a Poincar\'e invariant $M_4$.  Therefore, this limit
may provide an M-theory realisation of a four-dimensional field theory
with a zero cosmological constant and infinite Bose-Fermi mass
splitting \cite{witcos}, \ie these properties are a consequence
\cite{witcos} of the underlying ${\cal N}=1$ supersymmetry of the
three-dimensional field theory on $M_3$. (Note however, that in the
limit of large radius for the $S^1$, the size of the non-compact
manifold ${\cal M}_7$ of $G_2$ holonomy also grows, and so the
decoupling of the degrees of freedom associated with ${\cal M}_7$ from
those on the effective $M_4$ has to be addressed.  In fact for the
more general Spin(7) metrics (\ref{general2}) obtained in appendix A,
the presence of the additional non-trivial parameter $k$ allows us to
find a limit (\ref{decoupling}) 
where the $S^1$ and the ${\cal M}_7$ do fully decouple.
The manifold ${\cal M}_7$ can then be viewed as a blow-up of an orbifold
point in a compact manifold of $G_2$ holonomy, while the $S^1$
effectively decompactifies.)

    There are several differences between this wrapped D6-brane and
the D6-brane that comes from the $S^1$ reduction of the manifold $G_2$
holonomy with $S^3\times S^3$ principal orbits, which was discussed in
\cite{3wm}.\footnote{Some related ideas have also been discussed in
\cite{acha,gomi}.  See also the talk by E. Witten at the Santa Barbara
``David Fest,'' and forthcoming work by Atiyah and Witten.}  Since in
our case the radius of the $U(1)$ fibres becomes constant at infinity,
the D6-brane solution asymptotically approaches a product of $M_3$ and
the cone metric of the $S^2$ bundle over $S^4$.  The value of the
dilaton stabilises at large distance.  This situation is analogous to
the unwrapped D6-brane in the maximally-supersymmetric theory, where
it lifts to $D=11$ to become a product of $M_7$ with a
four-dimensional Taub-NUT.  On the other hand the dilaton becomes
singular at short distance, where the $U(1)$ fibres shrink to zero.
The D6-brane in \cite{3wm} has the opposite behaviour: the radius
diverges at large distance but stabilises to a fixed value at small
distance.

\subsection{M2-branes}

   Another application of metrics with Spin(7) holonomy is in the
construction of eleven-dimensional Lorentzian metrics that solve the
equations of eleven-dimensional supergravity theory, with the 4-form
$F_\4$ in (\ref{d11lag}) non-zero.  Metrics representing M2-branes are
given locally by
%%%%%
\be
ds^2 = 
H^{-{2 \over 3}} ( -dt^2 + dx_1^2 + dx_2 ^2 ) + H^{ 1 \over 3} ds_8^2\,, 
\label{brane}
\ee
%%%%% 
where $H$ is a harmonic function on the 8-manifold with metric
$ds_8^2$. Taking the metric $ds_8^2$ to be of holonomy Spin(7) guarantees
that the eleven-dimensional solution (including the 4-form $F_\4=dt
\wedge dx_1 \wedge dx_2 \wedge dH^{-1})$ admits at least one Killing
spinor. In the present case the simplest example to consider is when
$H$ depends only on the radial variable $r$.  For the case of $\bA_8$, with
metric (\ref{sol1}), one then has
%%%%%
\bea
H &=& 1+ Q\, \int^\infty_r { dx \over (x -\ell)^4 (x+3\ell)^2 }\nn\\
 &=&1 +\fft{Q(3r^3-3\ell\, r^2 -11\ell^2 r + 27\ell^3)}{192\ell^4\,
(r-\ell)^3(r+3\ell)} +
\fft{Q}{256\ell^5}\log\Big(\fft{r-\ell}{r+3\ell}\Big)\,.
\eea
%%%%%
In the case that $\ell$ and the constant are both positive, $H$ and will
be bounded and positive for $r>\ell$.  Near $r=\ell$ we have
%%%%%
\be
H  \propto (r-\ell)^{-3}.
\ee
%%%%%
This corresponds to the horizon of the M2-brane, which becomes
AdS$\4\times S^7$.  Thus we see that the M2-brane interpolates between
M$_3\times$ALC$_8$ at infinity and AdS$_4\times S^7$.
This solution represents an ${\cal N}=1$ dual supersymmetric gauge theory in
three dimensions that flows from the UV region (large distance) 
to the maximally supersymmetric conformal IR region (small distance).
  
   It is convenient to write
%%%%%
\be
H= { G\over (r-\ell)^3 } \,,
\ee
%%%%%
where $G$ is a positive smooth function for $r>a$, and $a$ is a
constant that is less than $\ell$.  Substitution in (\ref{brane}) shows that
the apparent singularity at $r=\ell$ is a coordinate singularity and
represents a degenerate event horizon. Near $r=\ell$, the metric tends
to the direct product AdS$_4$ times $\CP^3$, where AdS$_4$ is
four-dimensional anti-de-Sitter spacetime, \ie $SO(3,2)/SO(3,1)$ with its
standard Lorentzian metric.  It is possible to extend the Lorentzian
metric to $r<\ell$, but $r = -\ell$ represents a spacetime
singularity.  

  In the case that $\ell=-\td\ell$ is negative, the harmonic function
blows up near $r=3\td\ell$, and this appears to represent a spacetime
singularity. In particular it does not seem to be possible to
construct a spactime in which one pass between the positive and
negative $r$ regions along a smooth timelike (or indeed, as far as we
can see, spacelike) curve.  A similar analysis can be given for the more
general manifolds $\bB_8^\pm$ found in appendix A.

\subsection{Resolved M2-branes}

    Since both the Spin(7) manifolds $\bA_8$ and $\bB_8$ admit
$L^2$-normalisable harmonic 4-forms, we can construct resolved
M2-branes, whose metrics take the identical form as the regular
M2-brane (\ref{brane}), but with the 4-form $F_\4$ having an additional
contribution
%%%%%
\be
F_\4=dt\wedge dx_1\wedge dx_2\wedge dH^{-1} + m\, G_\4\,.
\label{4formbrane}
\ee
%%%%%
Instead of being harmonic, as in section 7.2, the function $H$ now satisfies
%%%%%
\be
\square H = -\fft{1}{48}m^2\, |G_\4|^2
\ee
%%%%%
on the Ricci-flat 8-dimensional space.

       For the manifold $\bA_8$, we have one harmonic normalisable
harmonic 4-form, and its magnitude is given in (\ref{g4square1anti}).
It follows that
%%%%%
\bea
H= 1 + \fft{m^2(3r^2 +26r +63)}{20(r+1)^2(r+3)^5}\,.
\label{h1}
\eea
%%%%% 
The solution is smooth everywhere; it interpolates between
eleven-dimensional Minkowski spacetime at small distance and
$M_3\times S^1\times {\cal M}_7$ at large distance.  Here ${\cal
M}_7$ is the 7-manifold of $G_2$ holonomy that is the $\R^3$ bundle
over $S^4$. 

       For the manifold $\bB_8$, we have two harmonic
normalisable 4-forms, whose magnitudes are given by
(\ref{g4square2anti}) and (\ref{g4square2self}).  It follows that the
function $H$ is given by
%%%%%
\be
H=1 + \fft{m^2(1323r^6 + 9786r^5 + 32937r^4 + 64428r^3 +
52237r^2 - 136934r + 29983)}{1680(r+1)^9(r-1)^2}\,,\label{h2}
\ee
%%%%%%%%
and
%%%%%
\be
H=1 + \fft{m^2(63r^4 - 80r^3 +114r^2+ 63)}{20(r+1)^7(r-1)^2}\,,
\ee
%%%%%%
respectively.

      The additional $G_\4$ term added to the 4-form field strength
(\ref{4formbrane}) has the possibility of breaking the supersymmetry of the
original unresolved brane solution.  The citerion for preserving  the
supersymmetry is that the covariantly-constant spinor $\eta$ in the
Ricci-flat 8-manifold should be such that \cite{hawtay,2beckers,clpres}
%%%%%
\be
G_{abcd}\, \Gamma^{bcd}\,\eta =0\,.
\ee
%%%%%
Using our results for the covariantly-constant spinor in $\bA_8$ or
$\bB_8$, we find that the supersymmetry will remain unbroken provided
that the functions $u_i$ in the harmonic 4-form (\ref{4formans})
satisfy precisely the linear relation given in (\ref{u123linear}).  Thus
our resolved M2-branes with anti-self-dual harmonic 4-forms in both the 
$\bA_8$ and $\bB_8$ manifolds are supersymmetric.  By contrast, the
resolved M2-brane using the self-dual harmonic 4-form in $\bB_8$ is
not supersymmetric.

        Resolved M2-branes in various manifolds were also constructed in
the previous papers \cite{clpres,cglp1,cglpcal}.  One important
difference is that in all three of the new resolved M2-brane solutions
obtained above, the 4-form $F_\4$ carries a magnetic M5-brane charge in
addition to the electric M2-brane charge.  The magnetic charge is given
by
%%%%%%%
\be
Q_m =\fft{1}{\omega_4}\int F_\4 = q\, m\,,
\ee
%%%%%
where $\omega_4$ is the volume of the unit 4-sphere, and $q=\ft12$, $\ft12$
and $\ft52$ for the three solutions respectively.  Thus our resolved
M2-brane solutions describe fractional magnetic M2-branes as wrapped
M5-branes, together with the usual electric M2-brane.  This
generalises the fractional D3-branes \cite{klebtsey,klebstra} of type
IIB theory to the case of M-theory.  It was argued in \cite{herkle}
that there should be no supersymmetric fractional M2-branes in
asymptotically conical manifolds.  Thus our fractional M2-branes do
not contradict the no-go theorem, since the $\bA_8$ and $\bB_8$
Spin(7) manifolds are not asymptocally conical, but instead have the
ALC structure with an $S^1$ whose radius tends to a constant at
infinity.

     In \cite{d2frac}, a supergravity solution of an ordinary D2-brane
together with a fractional D2-brane from the wrapping of a D4-brane
around the $S^2$ in a manifold of $G_2$ holonomy was obtained.  It was
conjectured \cite{cglpcal} that this D2-brane should be related to the
resolved M2-brane with a transverse 8-space of Spin(7) holonomy.  Here
we have provided a concrete realisation.  In our 2-brane solution, in
addition to the regular D2-brane (coming from the double dimensional
reduction of the M2-brane) and fractional D2-branes as wrapped D4-branes
(coming from the vertical reduction of the M5-brane), we also have wrapped
D6-brane charges.  This connection between D2-branes and M2-branes is
rather different from the one in a flat transverse space that was
discussed in \cite{imjy}, where the D2-brane was viewed as a periodic
array of M2-branes in the eleventh direction.

    All the M-theory solutions we discussed in this section can be
reduced on the principal orbits to give rise to four-dimensional domain
walls, given by
%%%%%%
\be
ds_4^2 = a^4\, b^2\, c^8 (H^{\ft53} (-dt^2 + dx_1^2 + dx_2^2) +
H^{\ft83}\, h^2\, dr^2)\,.\label{domainwall}
\ee
%%%%%

\section{Conclusions}

     In this paper, we have constructed new explicit complete
non-compact 8-metrics of Spin(7) holonomy.  Our procedure involved
writing down the ansatz (\ref{8ans}) for metrics of cohomogeneity one,
for which the principal orbits are $S^7$, described as a homogeneous
manifold with $S^4$ base and $S^3$ fibres that are themselves Hopf
fibred over $S^2$ and squashed along the $U(1)$.  This provides a more
general ansatz than the one that led to the previous complete
non-compact metric of Spin(7) holonomy obtained in
\cite{brysal,gibpagpop}.  We then showed that there exists a
first-order system of equations whose solutions yield Ricci-flat
metrics.  We first found simple solutions that give rise to two new
complete non-compact metrics.\footnote{It would be interesting also to
study these solutions using the methods developed in \cite{ootyas}.}
One is complete on the manifold that we denote by $\bA_8$, which is
topologically $\R^8$.  The other is complete on a manifold that we
denote by $\bB_8$, which is topologically the bundle of chiral spinors
over $S^4$.  Both the new metrics are asymptotically locally conical
(ALC), approaching $S^1\times {\cal M}_7$ locally at infinity, where
${\cal M}_7$ is the manifold of $G_2$ holonomy defined on the $\R^3$
bundle over $S^4$.  Thus the new manifolds have an asymptotically
cylindrical structure that is rather like Taub-NUT or Taub-BOLT.  This
is quite different from the asymptotically conical structure of the
previously-known Spin(7) example found in \cite{brysal,gibpagpop}.  At
short distance $\bA_8$ approaches Euclidean $\R^8$ locally, while
$\bB_8$ approaches $\R^4\times S^4$ locally.  We also obtained the
general solution to the first order equations (\ref{firstorder}) in
Appendix A, and showed that there exist further classes of regular
metrics of Spin(7) holonomy, complete on manifolds which we denote by
$\bB_8^\pm$, with the same topology as $\bB_8$.  These have a
non-trivial parameter $k$, with the earlier simple solutions
corresponding to the limit where $k=0$.

   We exhibited the Spin(7) holonomy of the new metrics by
constructing the covariantly-constant spinor associated with 
the special holonomy.  From this, we also constructed the calibrating
covariantly-constant self-dual 4-form $\Phi$.  We then showed that the
manifolds $\bA_8$ and $\bB_8$ both admit an $L^2$-normalisable 
anti-self-dual harmonic 4-form, and that $\bB_8$ also admits a second
$L^2$-normalisable harmonic 4-form, which is self-dual.

   The new Spin(7) manifolds have a variety of applications in
M-theory and string theory.  We discussed the eleven-dimensional
solutions obtained by taking the product of the Spin(7) metrics with
3-dimensional Minkowski spacetime, and the ten-dimensional solutions
obtained by reducing on the $U(1)$ fibres in the Spin(7) manifolds.
These give higher-dimensional analogues of the relation between
charged black holes in $D=4$ (Kaluza-Klein monopoles) and a product of
time and Taub-NUT in $D=5$.  In the limit where the string coupling is
strong, the world volume geometry $M_3\times S^1$ corresponds to the
large-radius limit of $S^1$, and thus it is effectively the Poincar\'e
invariant $M_4$.  This feature of M-theory compactified on these
Spin(7) manifolds may provide a concrete realisation of the proposal
in \cite{witcos} for explaining the vanishing cosmological constant,
the absence of Fermi-Bose mass degeneracy, and the absence of a 
massless dilaton in four-dimensional field theory.

   We also discussed M2-brane solutions, in which the $\bA_8$, $\bB_8$
or $\bB_8^\pm$ Spin(7) manifold replaces the usual flat 8-space
transverse to the membrane.  These solutions can be ``resolved'' by
adding an extra contribution to the 4-form in $D=11$, proportional to
a harmonic 4-form on the 8-manifold.  We showed that for each of the
$\bA_8$ and $\bB_8$ manifolds there is a resolved M2-brane solution
that preserves the single supersymmetry of the unresolved solution.
The second $L^2$-normalisable harmonic 4-form in the $\bB_8$ manifold
gives a resolved M2-brane that breaks supersymmetry.  The additional
contributions to $F_\4$ in the resolved solutions give rise to
non-vanishing magnetic M5-brane fluxes in the system, and hence our
solutions are the supergravity duals of fractional M2-branes.

\section*{Acknowledgements}

   We should like to thank Michael Atiyah, Andrew Dancer, Nigel
Hitchin, Dominic Joyce and Matthew Strassler for useful discussions.
H.L.~and C.N.P.~are grateful to UPenn for hospitality and financial
support during the course of this work.  M.C.~is supported in part by
DOE grant DE-FG02-95ER40893 and NATO grant 976951; H.L.~is supported
in full by DOE grant DE-FG02-95ER40899; C.N.P.~is supported in part by
DOE DE-FG03-95ER40917.  G.W.G.~acknowledges partial support from PPARC
through SPG\#613.

\newpage

\appendix
\section{General solution of the first-order equations}

   Here, we obtain expressions that yield the general solution
of the first-order equations (\ref{firstorder}).  Specifically, we show 
how the third-order equation for $f$ given in (\ref{feq}) may be solved.  
From this, one can then solve for $a$ and $b$ as in (\ref{absol}), and 
eliminate the spurious fourth constant of integration resulting from 
this procedure by substituting the results back into the first-order
equations (\ref{firstorder}).  Before presenting the general solution
of (\ref{feq}), we may note that it can be written in the
``factorised'' form
%%%%%
\be
f\, Q' - (f'+1)\, Q=0\,,\label{factored}
\ee
%%%%%
where $Q\equiv 2f\, W' + (f'-3)\, W$ and $W\equiv f'-1$.  In fact for
a generic solution, where $Q$ itself is non-zero, the solutions for
$a$ and $b$ can be written entirely algebraically in terms of $f$,
as
%%%%%
\be
a^2= \fft{(f'-1)(f'-3)\, f}{Q}\,,\qquad b^2 = \fft{2a^2}{(f'-1)^2}\,.
\label{absol2}
\ee
%%%%%
Thus for a solution where $Q\ne0$ the three integration constants for
the first-order system (\ref{firstorder}) are simply the three
integration constants for the third-order equation (\ref{feq}), and no
further substitution back into (\ref{feq}) is necessary.  As we shall
see below, $Q$ is non-vanishing for all but one degenerate solution of
(\ref{feq}).

    The first stage in solving (\ref{feq}) is to let
%%%%%
\be
f(r) = e^{\int^x g(s)\, ds}\,,
\ee
%%%%%
where the new radial variable $x$ is defined implicitly in terms of $r$ by
%%%%%
\be
\fft{df}{dr} = \fft1{x}\,.\label{frx}
\ee
%%%%%
    Using $f'$ to denote $df/dr$, we therefore have
%%%%%
\be
f' = \fft{1}{x}\,,\qquad
f''=-\fft{1}{f\, g\, x^3}\,,\qquad
f'''= \fft1{f^2}\, \Big[ \fft{3}{g^2\, x^5} + \fft1{g\, x^4} +
\fft{1}{g^3\, x^4}\, \fft{dg}{dx}\Big]\,.
\ee
%%%%%
Substituting into the original 3'rd-order equation (\ref{feq}) gives
the first-order equation
%%%%%
\be
2x\, \fft{dg}{dx} + 6 g + 6 x^2\, g^2 - x^2\,(x^2-1)(3x-1)\, 
g^3 =0\,.\label{geq}
\ee
%%%%%
Note that $f$ no longer appears explicitly; this is a
consequence of the scaling symmetry $f\longrightarrow \lambda\, f$,
$r\longrightarrow \lambda\, r$ of the original equation (\ref{feq}).
A further simplification can be achieved by setting  
$g=1/(\gamma(x)\, x^3)$, and also defining $x=1/\rho$.  Then, we find
%%%%%
\be
2 \gamma\, \fft{d\gamma}{d\rho} + 6\gamma = (1-\rho^2)(3-\rho)\,.
\label{gamma}
\ee
%%%%%

    We may first note that two specific solutions are
%%%%%
\be
\gamma=\ft12(1-\rho^2)\,,\qquad \gamma= \ft12(\rho-1)(\rho-3)\,.
\label{twosol}
\ee
%%%%%
The first of these leads back to our new solution in this paper.  The
second also gives a solution for $f$ (with an arbitrary multiplicative
constant of integration).  However, in this latter case it turns out
that after solving for $a$ and $b$ and plugging back into the original
first-order equations, this arbitrary constant of integration has to
be zero and so the second solution in (\ref{twosol}) is trivial.  In
fact it corresponds to solutions of (\ref{feq}) which, in the
factorised form (\ref{factored}), have $Q=0$.

   The next step in obtaining the general solution is to change
variable once again, from $\gamma$ to $z$, defined by
%%%%%
\be
z\equiv \fft{(1-\rho)^2}{2(1-\rho-\gamma)}\,.\label{zdef}
\ee
%%%%%
In terms of this new variable, (\ref{gamma}) becomes
%%%%%
\be
\fft{dz}{d\rho} = \fft{2z\, (1-z^2)}{\rho + 2z -1}\,.\label{zeq}
\ee
%%%%%
It turns out that the solution to this equation cannot be given
explicitly in the form of $z$ expressed as a function of $\rho$, but
it can be explicitly solved in the form of $\rho$ expressed as a
function of $z$.  To do this, it is convenient to characterise this
relation in the equivalent form
%%%%%
\be
u(z^2) + \fft{1-\rho}{2z}\, (1-z^2)^{1/4} = 0\,,\label{ueq}
\ee
%%%%%
for some function $u$ to be determined.

  Differentiating (\ref{ueq}) with respect to $\rho$, using (\ref{zeq}) to
substitute for $dz/d\rho$, and using (\ref{ueq}) itself to substitute for
$\rho$, we find that $u(y)$ satisfies
%%%%%
\be
4 y\, \fft{du(y)}{dy} + u(y) = (1-y)^{-3/4}\,,\label{ude}
\ee
%%%%%
where $y=z^2$.  The solution to this equation is
%%%%%
\be
u(y) = -\td k\, y^{-1/4} + {_2F_1}[\ft14,\ft34;\ft54;y]\,,
\ee
%%%%%
where $\td k$ is an arbitrary constant.
Thus we conclude that the general solution of (\ref{gamma}) for
$\gamma(\rho)$ is given by
%%%%%
\be
_2F_1[\ft14, \ft34;\ft54; z^2] = \fft{\td k}{\sqrt{z}} -
\fft{1-\rho}{2z}\, (1-z^2)^{1/4}\,,\label{gensol}
\ee
%%%%%
where $z$ is given by (\ref{zdef}) and $\td k$ is an arbitrary constant.

   The first special solution $\gamma=\ft12(1-\rho^2)$ in
(\ref{twosol}) corresponds to
$z=1$.  It is easily seen that this is indeed a special case of
(\ref{gensol}), with $\td k=[\Gamma(\ft14)]^2/(4\sqrt{\pi})$.  The second
special solution $\gamma=\ft12(\rho-1)(\rho-3)$ in (\ref{twosol}) 
corresponds to $z=-1$.  Again, this is seen to be a special case of
(\ref{gensol}), now with $\td k=\im\, [\Gamma(\ft14)]^2/(4\sqrt{\pi})$.

    Using identities for hypergeometric functions, another way to
write the general solution (\ref{gensol}) is
%%%%%
\be
(1-z^2)^{1/4}\, _2F_1[1,\ft12;\ft54; 1-z^2] = \fft{k}{\sqrt z}
                             +\fft{(1-\rho)}{2z}\, (1-z^2)^{1/4}\,.
\ee
%%%%%
(Here, the arbitrary constant $k$ is zero for the solutions with
$z=\pm1$.)  We may write the general solution as $\rho-1=v(z)$, where
%%%%%
\be
v(z) = \fft{2k\, \sqrt{z}}{(1-z^2)^{1/4}} -2z\,\, 
_2F_1[1,\ft12;\ft54; 1-z^2]\,.\label{vres}
\ee
%%%%%
Note that from (\ref{ude}) it follows that
%%%%%
\be
2z\, (1-z^2)\, \fft{dv}{dz} = v+ 2z\,.\label{vde}
\ee
%%%%%

   The general solution can now be presented explicitly, in the sense
that it is reduced to quadratures.  It is convenient in general to
take $z$ to be the radial coordinate in the metric.
Retracing the
steps of the various redefinitions, we eventually obtain
%%%%%
\be
c^2=f= \exp\Big[\int^z \fft{[v(z')+1]\,dz'}{v(z')\, (1-{z'}^2)}\Big]\,,
\qquad a^2 = \fft{[v(z)-2]\, z\, f}{v(z)\, (1+z)}\,,\qquad
b^2 = \fft{4 a^2}{v(z)^2}\,.\label{general}
\ee
%%%%%
The coordinate $r$ is given in terms of $z$ by
%%%%%
\be
dr=\fft{f\, dz}{v(z)\, (1-z^2)}\,.
\ee
Thus the general solution for the metric can be written as
%%%%%
\be
ds_8^2 = \fft{v\, f\, dz^2}{4z\, (1-z^2)(1-z)\, (v-2)} +
\fft{(v-2)\,z\, f}{(1+z)\, v}\, (D\mu^i)^2 + 
  \fft{4(v-2)\, z\, f}{(1+z)\, v^3}\, \sigma^2 + f\, d\Omega_4^2\,.
\label{general2}
\ee
%%%%%
Note that from (\ref{general}) we may express $f$ as
%%%%%
\be
f=\Big(\fft{1+z}{1-z}\Big)^{1/2}\, 
\exp\Big[\int^z \fft{dz'}{v(z')\, (1-{z'}^2)}\Big]\,.\label{fres2}
\ee
%%%%%

   Of the three expected constants of integration for the
first-order system (\ref{firstorder}) two are ``trivial,'' in the
sense that they correspond to a constant shift and rescaling of the
radial coordinate.  The non-trivial third constant of integration is
associated with $k$ in (\ref{vres}).

    In order to recognise the solutions that give rise to regular
metrics on complete manifolds, it is helpful to study the phase-plane
diagram for the first-order equation (\ref{vde}), which can be
expressed as
%%%%%
\be
\fft{dz}{d\tau} = 2z\,(1-z^2)\,,\qquad
\fft{dv}{d\tau} = v+2z\,,
\ee
%%%%%  
where $\tau$ is an auxiliary ``time'' parameter.   The solutions can
be studied by looking at the flows generated by the 2-vector field
$\{dz/d\tau,dv/d\tau\}=\{ 2z\,(1-z^2),  v+2z\}$ in the $(z,v)$ plane.  
For any such flow, it is then necessary to investigate the global
structure of the associated metric (\ref{general2}) for regularity.  
We find that
regular solutions can arise in the following four cases, namely
%%%%%
\bea
{\bf (1)} :&& z=1 \ \ \hbox{(fixed)};\qquad v=-2\ \ \hbox{to}\ \
v=-\infty\,,\nn\\
{\bf (2)} :&& z=1 \ \ \hbox{(fixed)};\qquad v=+2\ \ \hbox{to}\ \
v=+\infty\,,\nn\\
{\bf (3)} :&& z_0 \le z\le 1;\qquad v=+2\ \ \hbox{to}\ \
v=+\infty,\qquad (0<z_0 <1)\,,\nn\\
{\bf (4)} :&& 1\le z\le z_0;\qquad v=+2\ \ \hbox{to}\ \
v=+\infty,\qquad (1 <z_0 <\infty)\,.\label{sollist}
\eea
%%%%%
Note that $v=\pm\infty$ corresponds to the asymptotic large-distance
region, and in all four cases the metrics have similar asymptotic
structures, precisely as we have already seen in the $\bA_8$ and
$\bB_8$ cases.  $v=-2$ corresponds to the short-distance behaviour of
the $\bA_8$ metric, approaching Euclidean $\R^8$ at the origin where
the $S^7$ principal orbits degenerate to a point.  $v=2$ on the other
hand corresponds to the short-distance behaviour seen in the $\bB_8$
metric, approaching $\R^4\times S^4$ locally.  In fact solution (1) is
the metric (\ref{sol1}) on $\bA_8$ found in section 3, and solution
(2) is the metric (\ref{sol2}) on $\bB_8$ found there also.  These
both have $k=0$ in (\ref{vres}).

   Solution (3) arises when $k$ is any positive number, with $z_0$
being the corresponding value of $z$, with $0<z_0<1$, for which $v(z_0)=2$.
The value of $z_0$ is correlated with the value of $k$, ranging from
$z_0=0$ for $k=\infty$, to $z_0=1$ for $k=0$.\footnote{For the
case $k=0$, for which the regular solution is
$\gamma=\ft12(1-\rho^2)$ in (\ref{twosol}), and which leads 
to the metrics (\ref{sol1}) and (\ref{sol2}), the
quantity $z$ is not a good choice for the radial coordinate, since it
is fixed at $z=1$.  This case can be regarded as a singular limit
within the general formalism we are using here.  Specifically, if
we let $z=1-16 \ep^4\, \td\ell^4\, (r+\td\ell\, )^{-4}$, $k=2^{1/4}\, \ep$, and
choose the integration constant in (\ref{fres2}) so that 
$f=\ft12(r^2-\td\ell^2)$, then upon sending $\ep$ to zero we recover the
metric (\ref{sol2}).}  Near $z=1$ it follows
from (\ref{vres}) that we shall have
%%%%%
\be
v=2^{3/4}\, k\, (1-z)^{-1/4}  - 2 + \cdots\,,\qquad 
f=c_0\, (1-z)^{-1/2} + \cdots\,,
\ee
%%%%%
where $c_0$ is an arbitrary constant of integration.  Defining
$y\equiv (2c_0)^{-1/2}\, (1-z)^{-1/4}$, we see that as
$z\longrightarrow 1$ we shall have $y\longrightarrow\infty$ and
%%%%%
\be
ds_8^2 \approx dy^2 + \ft14 y^2\, (D\mu^i)^2 + \ft12 y^2\, d\Omega_4^2 + 
\fft{c_0}{k^2\, \sqrt2}\, \sigma^2\,,
\ee
%%%%%
and so this more general metric has the same large-distance
asymptotic form as  do $\bA_8$ 
and $\bB_8$.  Near $z=z_0$ we shall have $v(z) = 
2 + v'(z_0)\, (z-z_0) +\cdots$, and defining a new radial coordinate $R$ by
$(z-z_0)= \ft14 R^2$ near $z=z_0$, we shall have
%%%%%
\be
ds_8^2 \approx \fft{f_0}{2z_0\, (1-z_0^2)(1-z_0)\, v'(z_0)}\, 
\Big[ dR^2 + \ft14 v'(z_0)^2\, z_0^2\, (1-z_0)^2\, R^2\, 
[(D\mu^i)^2 + \sigma^2]\Big] + f_0\, d\Omega_4^2\,,\label{short3}
\ee
%%%%%
where $f_0$ is the value of $f$ at $z=z_0$.  From (\ref{vde}) we have
that $z_0(1-z_0)\, v'(z_0) =1$, and so we see from (\ref{short3}) that
at short distance the metric (\ref{short3}) approaches $\R^4\times
S^4$ locally.  Thus these more general solution (3) in (\ref{sollist})
with $k>0$ is complete on a manifold that is very similar to the
manifold $\bB_8$ of the solution (\ref{sol2}), with an $S^4$ BOLT at
$z=z_0$.  We shall denote it by $\bB_8^-$, where the superscript
indicates that $z$ starts from a value $z_0<1$ at short distance,
flowing to $z=1$ asymptotically.

   Solution (4) arises in the region where $z\ge1$, and again the flow
runs from an $S^4$  BOLT at $z_0$ (now $>1$) at which $v(z_0)=2$, to the
asymptotic region as $z$ approaches 1.  It follows from (\ref{vres}) that
in this case we should first introduce a new constant $\kappa$ such that
%%%%%
\be
v(z) = \fft{2\kappa\, \sqrt{z}}{(z^2-1)^{1/4}} -2z\,\, 
_2F_1[1,\ft12;\ft54; 1-z^2]\,.\label{vres2}
\ee
%%%%%n 
Since $v$ has the asymptotic form
%%%%%
\be
v\sim 2\kappa - \fft{2\sqrt{\pi}\, \Gamma(\ft54)}{\Gamma(\ft34)} +
\fft1{z} + {\cal O}(z^{-2})
\ee
%%%%%
at large $z$, one can show that we shall only be able to find the
required regular starting-point with $v(z_0)=2$ if $\kappa$ is bounded
by
%%%%%
\be
0 \le \kappa \le 1 + \fft{\sqrt{\pi}\, \Gamma(\ft54)}{\Gamma(\ft34)}\,.
\ee
%%%%%
(The lower limit corresponds to $z_0=1$, while the upper limit
corresponds to $z_0=\infty$.)
Under these circumstances we can find the necessary $z_0$ which
corresponds to an $S^4$ BOLT at short distance.  We shall denote this
solution by $\bB_8^+$.   Note that the simple solution $\bB_8$
in (\ref{sol2}) can be viewed as a $k\longrightarrow 0$ or
$\kappa\longrightarrow 0$ limit of the more complicated $\bB_8^-$ or
$\bB_8^+$ solutions.

   The arguments in section 4 show that in common with $\bA_8$ and
$\bB_8$ of section 3, the additional solutions $\bB_8^-$ and $\bB_8^+$ 
also have Spin(7) holonomy.

   We observed at the end of section 2 that a particular example of 
a solution of the first-order equations (\ref{firstorder}) is the 
direct product metric $ds_8^2=ds_7^2+d\varphi^2$, where $ds_7^2$ is the
Ricci-flat 7-metric of $G_2$ holonomy on the $\R^3$ bundle over $S^4$
\cite{brysal,gibpagpop}, and $\varphi$ is a coordinate on a circle.  We
are now in a position to see how this solution can arise as a limit of
our new Spin(7) metrics.  Specifically, it arises as the
$k\longrightarrow \infty$ limit of Solution (3) listed in
(\ref{sollist}).  This is the limit where the constant $z_0$, which sets
the lower limit for the range $z_0\le z\le 1$ for $z$, becomes zero.  

    At the same time as sending $k$ to infinity, we can rescale the
fibre coordinate $\varphi$ appearing the in definition
(\ref{kkvector}) for $\sigma=d\varphi + \cA_\1$, according to 
$\varphi\longrightarrow k\, \varphi$.  
From (\ref{vres}) and (\ref{fres2}) we see that
when $k$ becomes very large we shall have
%%%%%
\be
v\longrightarrow \fft{2k\, \sqrt{z}}{(1-z^2)^{1/4}}\,,\qquad
f\longrightarrow \Big(\fft{1+z}{1-z}\Big)^{1/2}\,,\label{decoupling}
\ee
%%%%%
and so in the limit of infinite $k$ the metric (\ref{general2})
becomes
%%%%%
\be
ds_8^2 = \fft{dz^2}{4z\, (1-z)^2\, (1-z^2)^{1/2}} 
+ \fft{z}{(1-z^2)^{1/2}}\, (D\mu^i)^2
+\Big(\fft{1+z}{1-z}\Big)^{1/2}\, d\Omega_4^2 + d\varphi^2\,.
\ee
%%%%%
Defining a new radial coordinate $r$ by $r^4=(1+z)\, (1-z)^{-1}$, we
see that this becomes $ds_8^2=ds_7^2 + d\varphi^2$, where 
%%%%%
\be
ds_7^2 = \fft{2dr^2}{1-r^{-4}} + \ft12 r^2(1-r^{-4})\, (D\mu^i)^2 
+ r^2\, d\Omega_4^2\,.
\ee
%%%%%
This can be recognised as the metric of $G_2$ holonomy on the manifold
${\cal M}_7$ of the $\R^3$ bundle over $S^4$, which was constructed in
\cite{brysal,gibpagpop}.\footnote{If we had not rescaled the fibre
coordinate $\varphi$ by a factor of $k$ before taking 
the limit $k\longrightarrow \infty$, the radius of the $S^1$
would have tended to zero.  This limit is known as the
Gromov-Hausdorff convergence \cite{joyce2}.}  Thus the family of new
Spin(7) manifolds that we are denoting by $\bB_8^-$ has a non-trivial
parameter $k$ such that the $k=\infty$ limit degenerates to ${\cal
M}_7\times S^1$, while the $k=0$ limit reduces to the case $\bB_8$
given by (\ref{sol2}).  

   Finally, we should stress that the analysis in this appendix
assumes that $f$ is not solely linearly dependent on $r$, since if it
is, we see from (\ref{frx}) that $x$ is then a constant.  This case is
easily analysed separately, and the conclusion is that the only
additional solution is the previous metric of Spin(7) holonomy found
in \cite{brysal,gibpagpop} (corresponding to $f=3r$, as we saw in
section 3).


\begin{thebibliography}{99}

\bm{brysal} R.L. Bryant and S. Salamon, {\sl On the construction of
some complete metrics with exceptional holonomy}, Duke Math. J. {\bf
58}, 829 (1989).

\bm{gibpagpop} G.W. Gibbons, D.N. Page and C.N. Pope, {\sl Einstein
metrics on $S^3$, $\R^3$ and $\R^4$ bundles}, Commun. Math. Phys.
{\bf 127}, 529 (1990).

\bm{clp0} M. Cveti\v{c}, H. L\"u and C.N. Pope,
{\sl Consistent warped-space Kaluza-Klein reductions, 
half-maximal gauged  supergravities and $\CP^n$ constructions},
Nucl.\ Phys.\ {\bf B597}, 172 (2001) [hep-th/0007109].

\bm{3wm} M.F. Atiyah, J. Maldacena and C. Vafa, 
 {\sl An M-theory flop as a large n duality}, hep-th/0011256.

\bm{cglpcal} M. Cveti\v{c}, G.W. Gibbons, H. L\"u and C.N. Pope,
{\sl Hyper-K\"ahler Calabi metrics, $L^2$ harmonic forms, 
resolved M2-branes,  and AdS$_4$/CFT$_3$ correspondence,}
hep-th/0102185.

\bm{cglp1} M. Cveti\v{c}, G.W. Gibbons, H. L\"u and C.N. Pope, {\it
Ricci-flat metrics, harmonic forms and brane resolutions}, hep-th/0012011. 

\bm{joyce} D. Joyce, {\sl A new construction of 8-manifolds with 
holonomy Spin(7)}, \\ math.DG/9910002.

\bm{bbmooy} K. Becker, M. Becker, D.R. Morrison, H. Ooguri, Y. Oz and
Z. Yin, {\sl Supersymmetric cycles in exceptional holonomy manifolds 
and Calabi-Yau  4-folds}, 
Nucl.\ Phys.\ B {\bf 480}, 225 (1996) [hep-th/9608116].

\bm{gibpap} G.W. Gibbons and G. Papadopoulos,
{\sl Calibrations and intersecting branes}, 
Commun.\ Math.\ Phys.\ {\bf 202}, 593 (1999)
[hep-th/9803163].

\bm{baisbate}  F. Bais and P. Batenburg, {\sl A new class of
higher-dimensional Kaluza-Klein monopole and instanton solutions}, 
Nucl. Phys. {\bf B253}, 162 (1985).

\bm{awacha}
A.A. Awad and A. Chamblin,
{\sl A bestiary of higher dimensional Taub-NUT-AdS spacetimes}, 
[hep-th/00012240]. 

\bm{GH}
G.W. Gibbons and S.W. Hawking, {\sl Classification of gravitational 
instanton symmetries}, Comm. Math. Phys. {\bf 66} (1979) 291. 

\bm{witcos} E. Witten,
{\sl Strong coupling and the cosmological constant}, 
Mod.\ Phys.\ Lett.\ A {\bf 10}, 2153 (1995) [hep-th/9506101].

\bm{acha} B.S. Acharya,  {\sl 
On realising N = 1 super Yang-Mills in M theory}, hep-th/0011089.

\bm{gomi} J. Gomis, {\sl D-branes, holonomy and M-theory},
hep-th/0103115.

\bm{hawtay} S.W. Hawking and M.M. Taylor-Robinson, {\it Bulk charges in
eleven dimensions}, Phys. Rev. {\bf D58} 025006
(1998) [hep-th/9711042].

\bm{2beckers} K. Becker and M. Becker, {\sl Compactifying M-theory to
four dimensions,} JHEP {\bf 0011}, 029 (2000) [hep-th/0010282].

\bm{clpres} M. Cveti\v{c}, H. L\"u and C.N. Pope, {\it Brane resolution
through transgression}, hep-th/0011023, to appear in Nucl. Phys. {\bf B}.

\bm{klebtsey} I.R. Klebanov and A.A. Tseytlin, {\sl Gravity duals of
supersymmetric $SU(N)\times SU(N+m)$ gauge theories},
Nucl. Phys. {\bf B578}, 123 (2000) [hep-th/0002159].

\bm{klebstra} I.R. Klebanov and M.J. Strassler, {\sl Supergravity and a
confining gauge theory: duality cascades and $\chi$SB-resolution of
naked singularities}, JHEP {\bf 0008}, 052 (2000), [hep-th/0007191].

\bm{herkle} C.P. Herzog and I.R. Klebanov, {\sl Gravity duals of
fractional branes in various dimensions}, hep-th/0101020.

\bm{d2frac}  M. Cveti\v{c}, G.W. Gibbons, H. L\"u and C.N. Pope,
{\sl Supersymmetric non-singular fractional D2-branes and NS-NS 2-branes,}
hep-th/0101096.

\bibitem{imjy} N. Itzhaki, J.M. Maldacena, J. Sonnenschein and
S.~Yankielowicz, {\sl Supergravity and the large $N$ limit of
theories with sixteen  supercharges,} Phys.\ Rev.\ D {\bf 58},
046004 (1998) [hep-th/9802042].

\bm{ootyas} T. Ootsuka and Y. Yasui, {\sl Spin(7) holonomy manifold
and superconnection}, Class. Quant. Grav. {\bf 18}, 807 (2001)
[hep-th/0010055].

\bm{joyce2} D. Joyce, private communication.

\end{thebibliography}
\end{document}